\documentstyle[psfig]{mn}
\def\H0{{\it H}$_0$}

\def\q0{{\it q}$_0$}
\def\kmps{km~s$^{-1}$}

\def\K{\rm K}
\def\ergps{erg~s$^{-1}$}

\def\nH{$N_{\rm H}$} 
\def\psqcm{cm$^{-2}$}
\def\ergpspsqcm{erg~cm$^{-2}$~s$^{-1}$}

\def\phpspsqcm{ph\thinspace s$^{-1}$\thinspace cm$^{-2}$}

\title[Iron K line complex in NGC\thinspace 1068]
{The iron K line complex in NGC\thinspace 1068: implications for X-ray reflection in the nucleus}
\author[K. Iwasawa, A.C. Fabian \& G. Matt]
{{K. Iwasawa$^1$, A.C. Fabian$^1$ and G. Matt$^2$}\\
\\
$^1$ Institute of Astronomy, Madingley Road, Cambridge CB3 0HA\\
$^2$ Dipartimento di Fisica, Universita di Roma III, Via della Vasca Navale 84, I-00146 Roma, Italy}
\date{}

\begin{document}

\maketitle

\begin{abstract}
We report a new analysis of ASCA data on the iron K line complex in
NGC\thinspace 1068. The line complex basically consists of three
components, as previously reported. A weak red wing of the 6.4 keV
fluoresence iron K line is found. A plausible explanation is Compton
scattering in optically thick, cold matter which can be identified with
an obscuring torus or cold gas in the host galaxy. We also show that this
`Compton shoulder' should be observable with ASCA using a simulated
reflection spectrum. In order to explain the two higher energy lines as
well as the cold 6.4 keV line, we fit the ASCA data with a composite
model of cold and warm reflection. This shows that cold reflection
dominates the observed X-ray emission above 4 keV; the estimated
scattering fraction from the warm medium is found to be $\sim$ 0.08 per
cent, an order of magnitude below the standard value of 1 per cent
obtained from previous observations in other wavebands. The two higher
energy lines have large equivalent width ($\sim 3$ keV) with respect to
the warm-scattered continuum, suggesting that efficient resonant
scattering operates. The line energies are systematically lower than
those expected from resonant lines for FeXXV and FeXXVI by $\sim 100$ eV.
The redshifts may be due to either the ionized gas of the warm mirror
receding at a radial velocity of 4000--5000 \kmps, or effects of Compton
scattering in a complicated geometry.
\end{abstract}


\section{INTRODUCTION}

Since the discovery of the polarized broad line region (Antonucci \& Miller
1985), the standard model for the Seyfert 2 galaxy NGC\thinspace 1068 is of 
a Seyfert 1 nucleus hidden
behind a thick obscuring torus. There is much observational support for this 
model (e.g., Antonucci 1993).
Contrary to the powerful nuclear activity observed at some other wavelengths,
the weakness of the observed X-ray luminosity and failure to detect any
direct component, even up to 20 keV, strongly suggests that
the direct X-ray radiation from the central source is blocked entirely 
from our direct view (Koyama et al 1989; Mulchaey et al 1995).

The lack of low energy absorption in the X-ray spectrum was interpreted
as evidence that the observed X-rays are the scattered radiation of a
hidden Seyfert 1 nucleus (e.g., Elvis \& Lawrence 1988). However,
extended soft X-ray emission ($\sim 13$ kpc) is resolved with the ROSAT
HRI (Wilson et al 1992), and an ASCA spectrum shows thermal emission,
probably coming from the spatially extended circumnuclear starburst,
dominates in the soft X-ray band (Ueno et al 1994). The ASCA spectrum
then no longer looks like a Seyfert 1, indicating very steep thermal
emission below 3 keV (which now agrees with the Einstein Observatory IPC
measurement by Monier \& Halpern 1984) and only above 3 keV, where a flat
continuum ($\Gamma\sim 1.2$) with strong iron K lines (Ueno et al 1994)
is seen, is emission presumed to originate from the active nucleus.

Based on the unification model, a large equivalent width (EW) 
of the iron K line was
predicted by Krolik \& Kallman (1987). This was later verified by
a Ginga observation ($EW\sim 1.3$ keV, Koyama et al 1989; Smith et al 1993).
The line centroid energy of the iron K line was $6.55\pm 0.1$ keV
which suggested a blend of cold and ionized lines, but it could not 
be resolved at the spectral resolution of the Ginga LAC 
(Turner et al 1989).
A subsequent BBXRT observation resolved the iron K line feature into three 
components for the first time (Marshall et al 1993). They were
identified with cold Fe (less ionized than FeXVII at 6.4 keV), FeXXV
and FeXXVI.

Several attempts to explain the iron K line feature have been made
(Band et al 1991; Marshall et al 1994), assuming that all 
X-ray emission is scattered light from the active nucleus and 
the scattering medium is optically thin in order to account for the lack of 
low energy absorption.
The observed continuum above 3 keV is unusually flat compared with
that in Seyfert 1 galaxies observed with ASCA 
($\Gamma\sim 1.9$; Nandra et al 1996; Reynolds 1996).
Together with the presence of a strong 6.4 keV line, 
this instead suggests that reflection from cold, thick matter is 
an important component, as observed in other Seyfert 2 galaxies such as
NGC\thinspace 6552 (Fukazawa et al 1994)
and the Circinus galaxy (Matt et al 1996a). To account for the three 
line components,
Matt, Brandt \& Fabian (1996) proposed a composite scenario in which 
two distinct reflections from an optically thick torus and 
an optically thin, warm scattering mirror together compose the observed 
X-ray spectrum of NGC\thinspace 1068.

We report here the discovery of a Compton shoulder on the 6.4 keV fluorescence 
iron K line and 
a spectral analysis of the ASCA PV data based on
the model by Matt et al (1996b).
We assume the distance of NGC\thinspace 1068 to be 22 Mpc throughout this
paper.

\section{THE ASCA DATA}

NGC\thinspace 1068 was observed on 1993 July 24 with ASCA (Tanaka, Inoue \& Holt 1994)
during the Performance Verification (PV) phase. 
Initial results have been published by Ueno et al (1994). 
We use only the Solid state Imaging Spectrometer (SIS; S0 and S1) data,
because of their superior spectral resolution for resolving the iron K
line complex.
The source data of 38 ks are extracted from a circular region of a 
radius 3.2 arcmin,
and the background data are taken from a source-free field in the same 
detector.
Spectral fits are performed with the S0 and S1 data
jointly while the figures presented
are made by summing both detectors for display purposes.
Errors on each spectral parameter are given at the 90 per cent confidence 
level for one interesting parameter.

\subsection{Compton shoulder}


\begin{figure}
\centerline{\psfig{figure=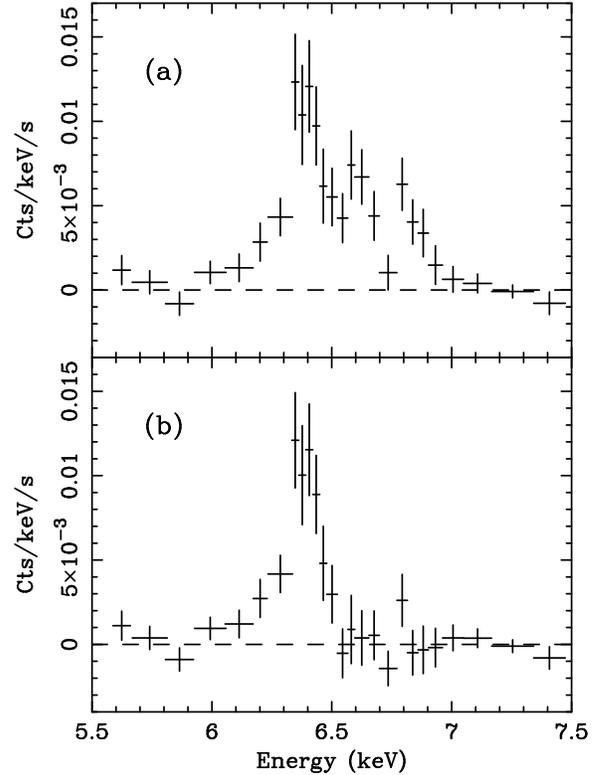,width=0.9\textwidth,angle=270}}
\caption{({\it a}) Total line profile of the iron K line complex in NGC\thinspace 1068
obtained from the ASCA SIS. The S0 and S1 are summed to display this figure. 
Note that the instrumental efficiency declines towards higher energy;
e.g., the efficiency at 6.9 keV is 15 per cent lower than at 6.4 keV.
({\it b}) The two higher energy lines are subtracted from the total line
profile. A clear line asymmetry is seen. The red wing can be identified with
a Compton shoulder of the 6.4 keV line.}
\end{figure}

\begin{table}
\begin{center}
\caption{Four gaussian fit to the iron K line complex. The line width of each
component is assumed to be $\sigma = 10$ eV, since no significant broadnening
($\sigma < 80$ eV) is found. The EWs are with respect to the single power-law
continuum with Galactic absorption (see text). Line-intensity ratios to
the 6.4 keV line (I/I$_{\rm 6.4}$) are also shown.}
\begin{tabular}{cccc}
E & I & EW & I/I$_{\rm 6.4}$\\
keV & $10^{-5}$\phpspsqcm & keV & \\[5pt]
$6.21\pm 0.16$ & $0.51^{+0.70}_{-0.50}$ & $0.11^{+0.15}_{-0.10}$ & 
$0.10^{+0.13}_{-0.09}$\\
$6.40\pm 0.03$ & $5.19^{+0.79}_{-1.27}$ & $1.10^{+0.17}_{-0.27}$ & 1.00 \\
$6.61\pm 0.03$ & $3.20^{+1.00}_{-0.80}$ & $0.69^{+0.22}_{-0.17}$ & 
$0.62^{+0.23}_{-0.18}$\\
$6.86\pm 0.05$ & $2.65^{+0.89}_{-0.87}$ & $0.58^{+0.19}_{-0.19}$ & 
$0.51^{+0.20}_{-0.20}$\\
\end{tabular}
\end{center}
\end{table}

\begin{figure}
\vspace{-3cm}
\centerline{\psfig{figure=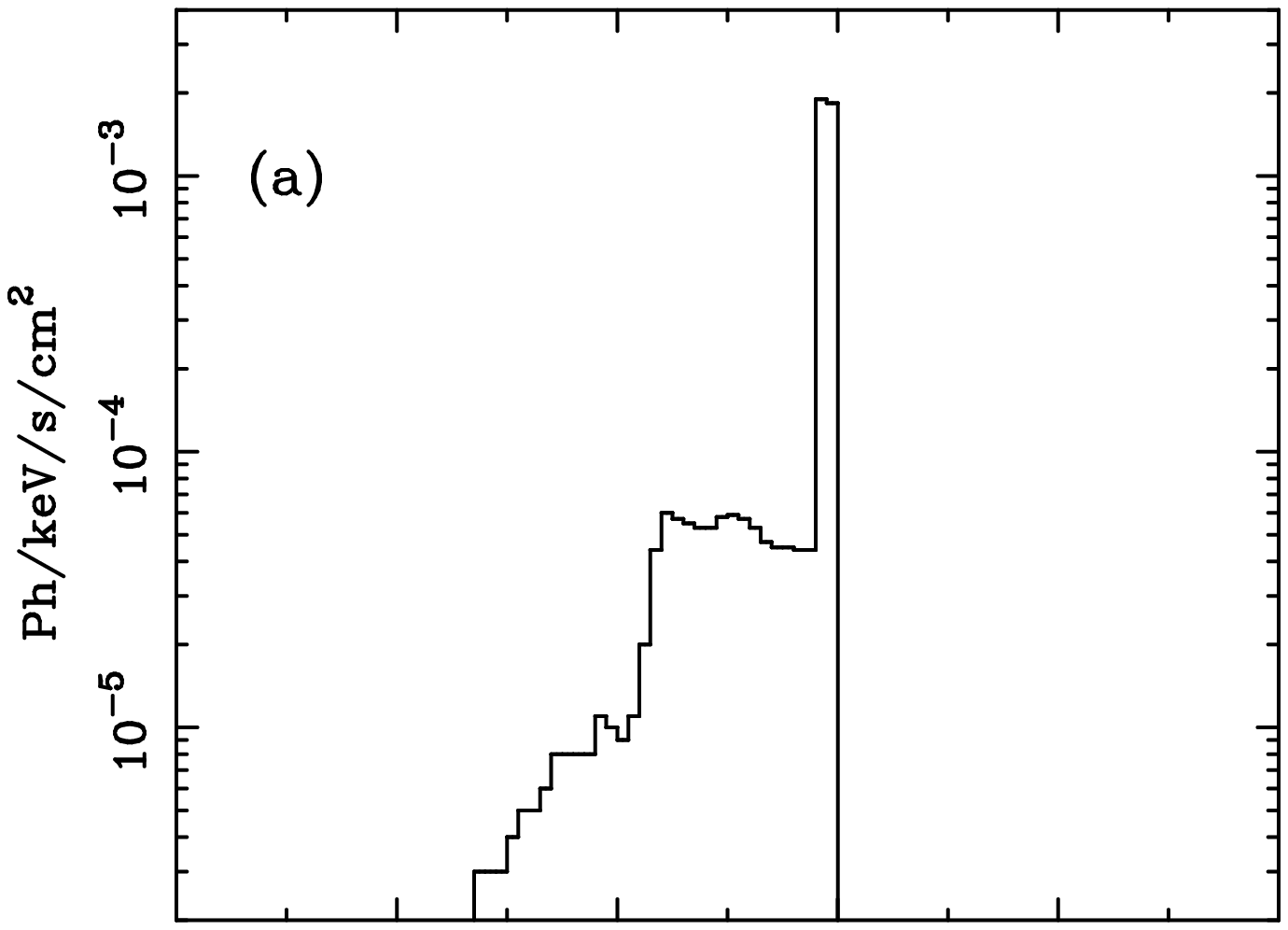,width=0.75\textwidth,angle=270}}
\vspace{-2.6cm}
\centerline{\psfig{figure=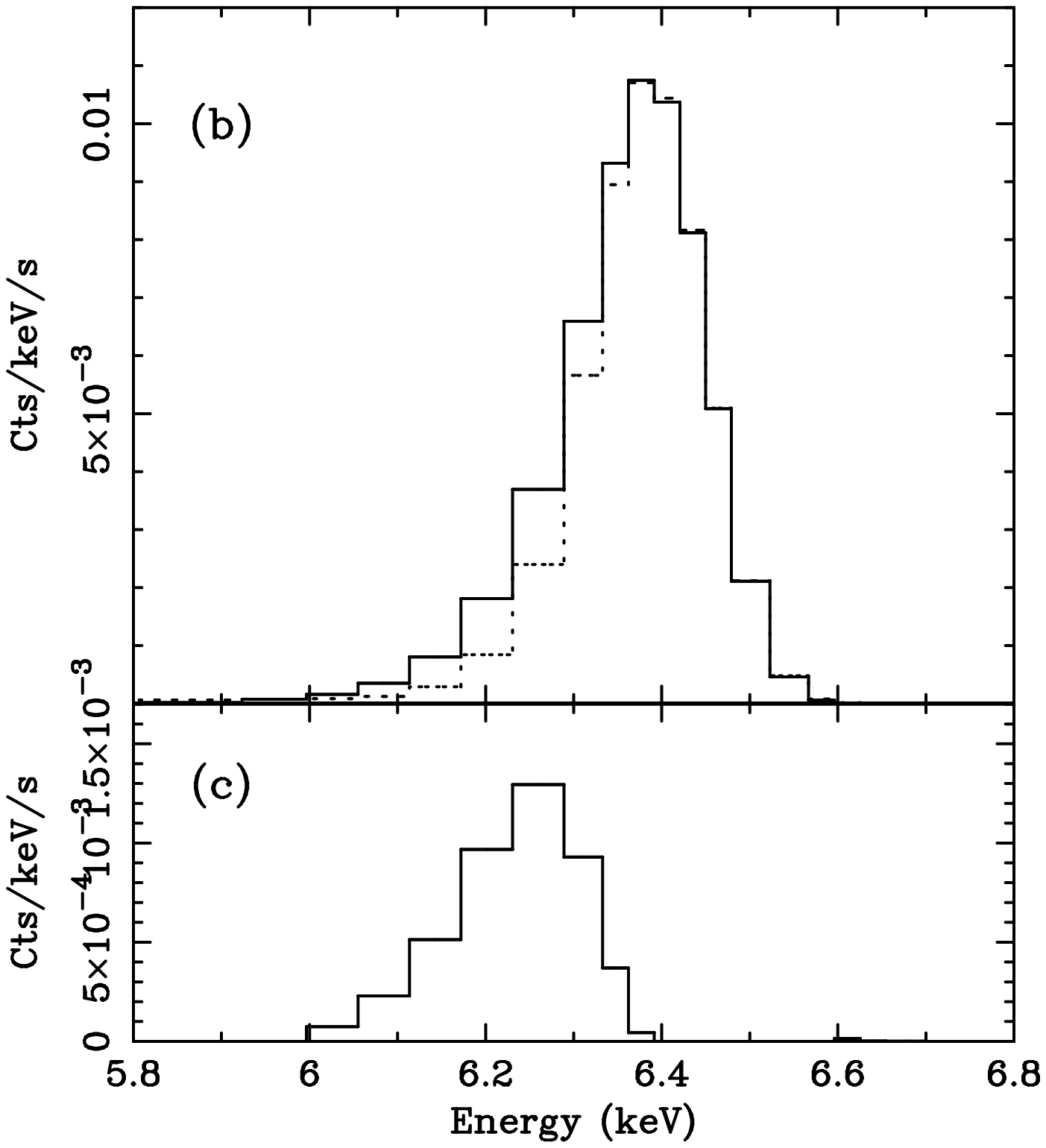,width=0.75\textwidth,angle=270}}
\caption{(a) Line profile of an iron K line from a Monte Carlo
simulation of reflection in thick, cold matter (see text). To show the
shape of the weak Compton shoulder, the intensity here has a logarithmic
scale. (b) The line profile (solid line) folded through the SIS response
from (a). A narrow ($\sigma = 10$ eV) gaussian with an intensity
normalized to the peak intensity of the line core is also shown by the
dotted line. Note that the spectral resolution of the SIS is FWHM
$\approx 130$ eV at 6.4 keV, appropriate for our data which were taken
during the PV phase. (c) Difference between the line with the Compton
shoulder and a narrow gaussian in (b). It is clear that the residual is
observable with the SIS, consistent with the observed result.}
\end{figure}

We show the observed line profile of the iron K complex in Fig. 1.
The underlying continuum is fitted with a single power-law model 
described below.
Since X-ray emission below 3 keV is dominated by starburst emission
(Ueno et al 1994) and there is evidence for emission lines due to
Ar and Ca (see below) between 3 and 4 keV, we use the 4--10 keV data.
Fitting the 4--10 keV data excluding the iron K line band 6--7 keV 
with a single power-law plus Galactic absorption \nH $= 4.5\times 10^{20}$
\psqcm 
~(Stark et al 1992), leaves the total line profile shown in Fig 1a. 
The photon-index is $\Gamma = 0.35_{-0.30}^{+0.66}$, flatter than that 
given in Ueno et al (1994) because of the exclusion of the 3--4 keV 
band where some emission lines are present, causing the slope to steepen.

As previously suggested by Marshall et al (1993) and Ueno et al (1994),
three emission-line peaks are seen. These three lines do not show
evidence for significant broadening ($\sigma < 80$ eV). However, there is
a weak wing on the red side of the 6.4 keV line. To clarify this, we show
only the 6.4 keV line profile in Fig. 1b, which was produced by fitting
the two higher energy lines with a narrow ($\sigma =10$ eV) gaussian and
subtracting them from the total line profile. The subtraction of the two
lines is fairly good as there are no significant residuals seen above 6.5
keV. The line profile is clearly more complicated than a single gaussian;
the core emission is consistent with the SIS energy resolution, whereas a
wing can be seen on the red side in contrast. Fitting the red wing by a
gaussian gives a line energy of $6.21\pm 0.16$ keV and an intensity one
tenth of the 6.4 keV line with $\sim 90$ per cent confidence level. The
four gaussians plus a power-law model now
gives a good fit to the 4--10 keV data with $\chi^2 = 54.98$ for 70
degrees of freedom. The line energies, equivalent widths and relative
line intensities of each line component are summarized in Table 1.

The possibility of a symmetric broad base to the 6.4 keV line cannot be
excluded, since the subtraction of the higher energy lines will have
eliminated any blue wing. A double gaussian, with a narrow and a broad
component, fit to the 6.4~keV line profile is slightly worse than the one
with the red wing (by $\Delta\chi^2\simeq 1.4$), but acceptable. In this
case, the line width of the broad component is $\sigma\sim 0.3$ keV.

Compton scattering is a likely mechanism to explain the red wing. A line
extending only to lower energy indicates that the electrons responsible
for scattering are cold (Pozdnyakov, Sobol \& Sunyaev 1979). Theoretical
calculations of Compton reflection spectrum from optically thick, cold
matter have already predicted such a weak red wing, the so called
`Compton shoulder' (George \& Fabian 1991; Reynolds et al 1994), which is
a characteristic feature due to Compton scattering of the 6.4 keV
fluorescent line photons. It consists of a series of shoulders extending
towards low energies. Computed line shapes due to the Compton scattering
are presented in Hatchett \& Weaver (1977); Pozdnyakov et al (1979);
George
\& Fabian (1991); and Sunyaev \& Churazov (1996). The first shoulder is due
to single scattering and is spread over two Compton wavelengths ($\simeq
160$ eV); multiple scattering makes a further weak low energy tail.

If pure reflection from thick, cold matter is observed with ASCA, the
Compton shoulder should be resolvable with the SIS. To demonstrate this
we have used a Monte Carlo code kindly provided by C. Reynolds (see also
George \& Fabian 1991). The line profile derived from a Monte Carlo
reflection spectrum with spectral resolution of 10 eV (Fig. 2a) is
compared, after folding through the SIS response, with that of a narrow
($\sigma = 10$ eV) gaussian (Fig. 2b). The difference bewtween the two
profiles should be detectable (Fig. 2c). The shape of the Compton
shoulder depends slightly on geometry (e.g. Matt, Fabian \& Reynolds
1996) and its height also depends on the abundance ratio of iron and the
lighter elements, mainly oxygen, which absorb the 6.4 keV photons (George
\& Fabian 1991). In order that the shoulder is maximally broad, a large
back-scattered component is required. The quality of the present data is
not good enough for detailed spectral modelling, however, and we merely
assume that reflection occurs from a face-on, semi-infinite slab with an
optical depth of unity. The strength of the simulation has been adjusted
to the observed intensity level. Although the residual profile due to the
Compton shoulder shows some asymmetry, fitting this with a gaussian
gives the line centroid energy $\sim 6.24$ keV and the intensity ratio to
the gaussian line-core $\sim 12$ per cent. These values are in good agreement
with the gaussian modelling of the observed line profile (see Table 1).

Note that the broad line from a relativistic accretion disk, such as that
observed in Seyfert 1 galaxies (e.g., Tanaka et al 1995), is an unlikely
explanation for the red wing, since we do not directly see the
relativistic line emitting region (which emerges from within $\sim40$
gravitational radii). A scattered relativistic line is also implausible.
The scattering cone appears to have a half opening angle of $\sim
40^{\circ}$ (Evans et al 1991; Tsvetanov et al 1996) and the inclination
angle of the accretion disk in NGC1068 with respect to most of the
scattering volume must therefore be near $40^{\circ}$. In this geometry
the red wing of the diskline would be skewed down to 4 keV and peaked at
energies well below 6 keV (see inclination angle dependence of diskline
profiles in Fabian et al 1989) giving rise to a much broader red wing
than observed.

\subsection{Spectral fit with the composite model}


\begin{figure}
\vspace{-0.8cm}
\centerline{\psfig{figure=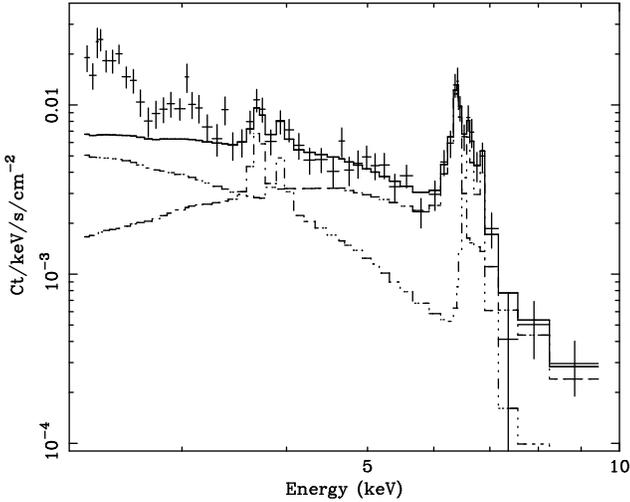,width=0.7\textwidth,angle=270}}
\caption{ASCA SIS spectrum of NGC\thinspace 1068 fitted with the composite model.
The best-fit model for the 3.2--10 keV band is shown as well as the cold
reflection and warm scattered continuum plus ionized lines. The data
below 3.2 keV down to 2.3 keV are also shown so that excess emission due
to the starburst can be seen.}
\end{figure}

\begin{figure}
\vspace{-0.8cm}
\centerline{\psfig{figure=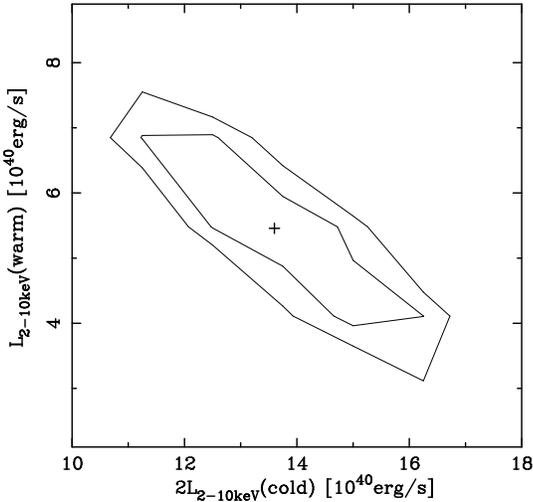,width=0.75\textwidth,angle=270}}
\vspace{-0.8cm}
\caption{Contour plot of the 2--10 keV luminosities of the cold and 
the warm reflection continua obtained from the composite model fit. 
The luminosities are derived assuming the distance of NGC\thinspace 1068 to be
22 Mpc. Note that the cold reflection luminosity is integrated over 
$2\pi$ (see text). 
Contours are drawn at 68 and 90 per cent confidence levels for 
two interesting parameters. The best-fit values, $2L_{\rm C}
=1.36\times 10^{41}$\ergps ~and $L_{\rm W}=
5.47\times 10^{40}$\ergps, are shown by a plus symbol. }
\end{figure}

The presence of the Compton shoulder indicates that reflection from
optically thick cold matter is an important process in producing the
observed hard X-ray emission in NGC\thinspace 1068. However, pure cold
reflection alone does not explain the two higher energy lines.

It is likely that the higher energy lines are produced in optically 
thin ionized (warm) matter which is located above the obscuring
torus and electron-scatters the continuum of 
the central source, as originally proposed by Krolik \& Kallman (1987).
If the ionized scattering medium is optically thin enough,
photoelectric absorption has little effect on
the electron-scattered continuum so the emergent spectrum retains 
almost all the original shape of the central source, in the ASCA band.

Thus, as proposed by Matt et al (1996b),
a combination of cold and warm reflections could
be a plausible description for the observed hard X-ray emission from 
NGC\thinspace 1068.
We therefore fit the ASCA spectrum with the composite model
described above.

We assume that the incident spectrum of the central source is a
power-law with $\Gamma = 2.0$, typical of Seyfert 1s 
(e.g., Nandra et al 1996; Reynolds 1996).
The pure cold reflection continuum and the warm-scattered continuum
are approximated by the viewing-angle averaged Lightman \& White (1988) type 
reflection model and a power-law of $\Gamma = 2.0$ with 
no intrinsic absorption, respectively.
The iron K line complex is modelled by four gaussians including
the Compton shoulder.
Weak line features due to calcium between 3.6--4 keV are also 
modelled by narrow gaussians.
The 3.2--10 keV data are fitted with the composite model, and 
the best fit model gives $\chi^2 = 70.14$ for 83 degrees of freedom.
The EW of each line are shown in Table 2.
Note that the EW of the cold iron K line is the sum of the 6.4 keV
line-core and the Compton shoulder measured 
with respect to the cold reflection continuum,
and those of the two higher energy lines are with respect to the warm-scattered
$\Gamma = 2.0$ power-law continuum.

Although only the 3.2--10 keV data are fitted, to show excess emission
due to the starburst below the low energy bound, the 2.3--10 keV data are
shown in Fig. 3. Two significant lines at $3.71\pm 0.04$ keV and $3.94\pm
0.09$ keV are identified with a neutral and helium-like calcium (CaXIX).
The EW of the neutral Ca line with respect to the cold reflection
continuum is $EW=190\pm 110$ eV, which is similar to that in
NGC\thinspace 6552 (Fukazawa et al 1994) but larger than that expected
from a Monte Carlo simulation ($\sim 50$ eV; Reynolds et al 1994). This
is further evidence for cold reflection. Another line feature around 3.1
keV is due to $K\alpha$ emission from ArXVII which is probably from both
the thermal emission and the warm scattering matter.

\begin{table}
\begin{center}
\caption{Equivalent widths of two calcium-K and three iron-K lines 
respect to each continuum conponent obtained from the composite model fit.
EW$_{\rm C}$ and EW$_{\rm W}$ are equivalent widths respect to the 
cold reflection continuum and the power-law warm-reflection continuum,
respectively. Note that 
the EW for the 6.4 keV line is the sum of the line-core 
and Compton shoulder.}
\begin{tabular}{ccc}
E$_{\rm line}$ & EW$_{\rm C}$ & EW$_{\rm W}$ \\
keV & keV & keV \\[5pt]
3.71 & $0.19^{+0.11}_{-0.11}$ & --- \\
3.94 & --- & $0.11^{+0.06}_{-0.10}$ \\
6.40 & $1.21^{+0.26}_{-0.28}$ & --- \\
6.61 & --- & $3.49^{+1.09}_{-0.87}$ \\
6.86 & --- & $2.92^{+0.98}_{-0.96}$ \\
\end{tabular}
\end{center}
\end{table}

The 2--10 keV fluxes of the two components estimated from the best-fit
model are $F_{\rm X}({\rm cold})=2.5\times 10^{-12}$\ergpspsqcm ~for 
the cold reflection,
and $F_{\rm X}({\rm warm})=1.0\times 10^{-12}$\ergpspsqcm 
~for the warm scattered 
continuum. Confidence contours of the corresponding luminosities are given
in Fig. 4, using a distance of 22 Mpc.
The best-fit values in the 2--10 keV band 
are $6.8\times 10^{40}$\ergps 
~for the cold reflection continuum
and $5.5\times 10^{40}$\ergps 
~for the warm reflection continuum, respectively.
It should be noted that the luminosity is integrated 
over $2\pi$ for the cold reflection because of the assumed geometry 
whilst over $4\pi$ for the warm reflection.
However, for a direct comparison of observed intensity between the 
two components, the cold reflection luminosity multiplied by 2 is 
shown in Fig. 4.

If significantly large absorption (say, \nH $>10^{22}$\psqcm) or a
starburst component is included in the fit, the value of $L_{\rm W}$
would be reduced slightly. Fitting the data in a higher energy band (e.g.
4--10 keV) gives poorer constraints on $L_{\rm C}$ and $L_{\rm W}$.
Results are consistent with those in Fig. 4, although the warm reflection
component is not necessary at the 90 per cent confidence level.

\section{DISCUSSION}

\subsection{Intrinsic luminosity}
A rough estimate of the intrinsic X-ray luminosity can be obtained using
the warm scattering luminosity $L_{\rm W}\approx 5.5 \times
10^{40}$\ergps ~derived from the composite model fit. For highly-ionized
iron such as FeXXV and FeXXVI, resonant scattering plays an important
role in making the EW large, if the scattering medium is optically thin
(Band et al 1991; Matt et al 1996), but it is easily suppressed below the
large observationally inferred values (Table 2) when the column density
is, say, larger than $10^{22}$\psqcm ~(e.g., see Matt et al 1996b). This
indicates that the optical depth of the warm scattering mirror is small
enough to allow efficient resonant scattering. Thus assuming a scattering
depth ($nl\sigma_{\rm T}$) of 0.002$\tau_{0.002}$ so that the column
density of the scattering mirror is \nH$\sim 3\times
10^{21}\tau_{0.002}$\psqcm, and the fraction of the sky occupied by the
scattering mirror, $0.25(\Omega/4\pi)_{0.25}$, the intrinsic luminosity
is estimated to be $L' = L_{\rm W}\tau^{-1}(\Omega/4\pi)^{-1}
\simeq 1\times 10^{44}\tau_{0.002}^{-1}(\Omega/4\pi)_{0.25}^{-1}$\ergps
~in the 2--10 keV band, where
the density and depth of the scattering medium are $n$ and $l$, 
respectively, and 
$L_{\rm W}$ is the measured luminosity of the warm reflection continuum.
This estimate is entirely independent of data in other wavebands.

There are further possible ways to estimate the intrinsic luminosity.
Using the correlation between X-ray and [OIII]$\lambda 5007$ luminosities
(log$L_{\rm [OIII]}/{\rm log}L_{\rm 2-10keV} \sim -1.8$) derived by
Mulchaey et al (1995), the 2--10 keV luminosity would be $L'\simeq
7\times 10^{43}$\ergps. Note that the [OIII] flux was measured with a
large aparture (10--15 arcsec, see Whittle 1992) containing the strong
starburst region which could contribute some fraction of the flux. The
estimate from the above, X-ray based, argument is consistent with this
value. For a comparison, the far infrared luminosity measured with IRAS
is $5\times 10^{44}$\ergps (e.g., David, Jones \& Forman 1993), although
this does contain a significant contribution from the circumnuclear
starburst.

\subsection{Reflection from cold matter}

The detection of the Compton shoulder of the 6.4 keV iron fluorescent
line is strong evidence that we are seeing reflection from optically
thick, cold matter. The presence of the neutral calcium line supports
this. The flat 4--10 keV continuum ($\Gamma\sim 0.4$) suggests that cold
reflection is a major component of the observed X-ray emission from the
active nucleus which dominates the observed radiation only in the energy
band above 3 keV. In fact, the fit with the composite model of cold and
warm reflections (Section 2.2) indicates that $\sim 71$ per cent of the
total 2--10 keV flux comes from cold reflection.

Presumably the cold reflection occurs at the inner surface 
of the optically-thick, obscuring torus which entirely blocks
the direct radiation from the central source in our direction (Ghisellini,
Haardt \& Matt 1993; Matt et al 1996b).
For such a geometry only part of the reflection surface is visible to us.
If that fraction is $f_{\rm C}$, the observed luminosity is described as
$L_{\rm C} = f_{\rm C}\eta L'$, 
where $\eta$ is the total albedo of a cold reflector
subtending a solid angle of $2\pi$.
Using values of $L_{\rm C} = 6.8\times 10^{40}$\ergps (see Section 2.2
and Fig. 4), $\eta = 0.022$ (in the 2--10 keV band), and 
$L' = 7\times 10^{43}L'_{43.8}$\ergps ~(see Section 3.1), 
$f_{\rm C}$ is then found to be 
$\approx 4.4\times 10^{-2}L^{\prime -1}_{43.8}$.

The small value for $f_{\rm C}$ suggests a high inclination of the obscuring
torus (close to edge-on) if its inner surface is the reflecting source.
Alternatively a torus of a small radial/axial ratio (Pier \& Krolik 1993)
is also possible, if it is not nearly edge-on.
Recent water maser observations by Gallimore et al (1996) and 
Greenhill et al (1996)
suggest a high inclination $>82$ degrees of the maser disk which
is on a pc scale with a toroidal geometry and
is identified with the X-ray absorbing gas torus
at large optical depth (i.e., \nH$>10^{25}$\psqcm, Koyama et al 1989).
A rather broad [OIII]$\lambda 5007$ emission-line cone (projected 
opening angle $\theta_{\rm proj}\sim$ 65--80$^{\circ}$, Cecil, Bland 
\& Tully 1990; Evans et al 1991) also favours an edge-on torus over one
with a large scale-height.

Another possible cold reflector is the cold gas in the bulge and
disk of the host galaxy (Fabian 1977).
Strong 6.4 keV line emission is observed from molecular clouds 
in our Galactic centre, where a strong 
iron K line emitter is Sgr B2, $\sim 100$ pc away from the Galactic
centre. The X-ray spectrum there also shows a complex of iron K lines
(Koyama et al 1996).
As the stellar disk of NGC\thinspace 1068 is almost face-on, contrary to the 
torus, radiation from the central source escaping from the torus 
can pass through some of the cold gas clouds in the host galaxy.
The likely distances of cold clouds from the active nucleus and
the disk-like geometry make the covering factor small,
and consistent with the small $f_{\rm C}$.
If this is the case, future high spatial resolution mission like
AXAF would resolve a conical shape to the iron K line emission.

The cold reflection spectrum has been studied by many authors
(Guilbert \& Rees 1988; Lightman \& White 1988; George \& Fabian 1991; 
Matt, Perola \& Piro 1991; Reynolds et al 1994).
The cold reflection-dominated spectrum in NGC\thinspace 1068 predicts that
a spectral break due to Compton down-scattering would be observed 
around 30 keV. This is consistent with the CGRO/OSSE upper limit
and may be observable with high quality data from 
RXTE and SAX.

\subsection{Reflection from warm matter}

Cold reflection is a unique phenomenon at X-ray wavelengths.
Warm scattering is much more relevant to the observed optical/UV scattering.
The 2--10 keV luminosity of the warm-scattered continuum is
$L_{\rm W}\approx 5.5\times 10^{40}$\ergps, 
which implies a scattering fraction
in the warm matter to be $f_{\rm W}= L_{\rm W}/L'\simeq 7.8\times 10^{-4}
L_{43.8}^{\prime -1}$.
Various other observations of NGC\thinspace 1068 suggest that the scattering fraction is 
approximately 1 per cent (summarized in Pier et al 1994).
The X-ray scattering fraction appears to be an order of magnitude lower.
Pier et al (1994) deduced the bolometric luminosity of NGC\thinspace 1068 to be
$\sim 7\times 10^{44}f_{0.01}^{-1}$\ergps, where the scattering fraction 
is 0.01\thinspace $f_{0.01}$.
Since they assumed that the whole hard X-ray emission 
above 2 keV (as observed with BBXRT,
Marshall et al 1993) is scattered by the warm mirror,
contrary to our result, their estimate has been modified slightly.

If the two higher energy lines are due to FeXXV and FeXXVI, the
ionization parameter of the warm scattering matter, $\xi=L'/nR^2$, is
about several hundreds (Kallman \& McCray 1982; Krolik \& Kallman 1987;
Band et al 1991), where $R$ is a distance of the scattering medium from
the central source. Assuming $l=R$, the column density can be written as
$nl=L'(\xi l)^{-1}$. Since $f_{\rm W}=\sigma_{\rm T}nl(\Omega/4\pi)$, $$l
= {\sigma_{\rm T}L'^{2}\over{\xi L_{\rm W}}}({\Omega\over{4\pi}}).$$ If
$\xi = 600 \thinspace\xi_{\rm 600}$ and $\Omega/4\pi = 0.25
(\Omega/4\pi)_{\rm 0.25}$ (e.g., Evans et al 1991) are taken, then the
size of the scattering region is approximately $l \simeq
8\thinspace\xi_{\rm 600}^{-1}L_{43.8}^{\prime 2}(\Omega/4\pi)_{\rm 0.25}$
pc. This is large enough to be visible to us without occultation by a
pc-scale torus; direct X-ray imaging of this is impossible at present as
the angular size is the order of 0.1 arcsec ($\approx$ 11 pc at D = 22
Mpc). The size of the optical/UV scattering nebula observed by
spectropolarimetry with the Hubble Space Telescope (HST) is several tens
pc or up to 200 pc (Antonucci et al 1994; Capetti et al 1995), the warm
X-ray mirror is therefore smaller. The column density $nl = 5\times
10^{21}L_{43.8}^{\prime -1}(\Omega/4\pi)_{0.25}$\psqcm, which is small
enough to allow efficient production of the resonant scattering lines of
FeXXV and FeXXVI as mentioned in Section 3.1. To maintain the high value
of $\xi$ in order to emit the highly ionized iron lines over a large
scale, a low mean electron density such as $\sim 200$ cm$^{-3}$ is
required.

\subsection{Complexities}

We have discussed X-ray scattering from a warm medium in the context of
the standard unification model as proposed by Krolik \& Kallman (1987)
and Matt et al (1996b), i.e., the scattering medium is located outside 
the obscuring torus and is directly visible to us.
However, there are some problems: (1) a very small scattering fraction
$f_{\rm W}\sim 0.8\times 10^{-3}$; (2) disagreement of line energies
discussed below; and (3) the location of the region where the
higher energy lines are formed.

We note that the line energies for the two higher energy lines (6.61 keV
and 6.86 keV) disagree with the energies of the resonant lines of FeXXV,
6.70 keV, and FeXXVI, 6.97 keV, (see Table 1) whilst that of the cold
iron K line agrees with 6.40 keV perfectly. The shift of the line
energies are $-90\pm 30$ eV for FeXXV and $-110\pm 50$ eV for FeXXVI. We
have checked that there are no significantly strong iron lines at the
observed energies from any other ionization states (e.g., Beiersdorfer et
al 1993). Interestingly, the shift of the two line energies are similar
to each other ($\sim 1.5\pm 0.6$ per cent lower than rest energies). If
the redshift is due to the Doppler effect from receding warm scattering
matter, then the radial velocity is $\sim$ 4000--5000 \kmps. The optical
polarized broad H$\beta$ line is also redshifted but only by 600
\kmps~(Antonucci \& Miller 1985) from the narrow line component, which is
explained as scattering in outflowing wind (Krolik \& Begelman 1986;
Balsara \& Krolik 1993). As the discrepancy between the velocities is too
large, the same outflowing wind is perhaps not an explanation. It should
be noted however that the X-ray lines are produced in the mirror itself
whilst the polarized broad line is scattered by the mirror.

The location of the gas emitting the two higher energy lines is an issue.
It is supposed to lie outside the torus. Systematic studies of Seyfert 1
galaxies with ASCA show that a large fraction of them show evidence for
partially ionized gas in the line of sight, implied by absorption edge
feature mainly due to oxygen, OVII and OVIII (Reynolds 1996). A recent
study of the warm absorber in MCG--6-30-15 (Otani et al 1996) suggests
that the OVII and OVIII absorbers occupy distinct regions; the
variability study locates the OVII absorber $>1$ pc far from the central
source, and the OVIII absorber in the broad-line region. If we observed
this object from a different viewing angle, the OVII warm absorber would
be seen as a scattering mirror. However, the ionization state implied
from OVII ($\xi\leq$ a few tens erg cm s$^{-1}$) is too low for the
emission of FeXXV and FeXXVI. If ionized gas in the nucleus of
NGC\thinspace 1068 has a similar ionization structure to that of
MCG--6-30-15, the gas emitting the highly-ionized iron line should be
located well within the torus like the OVIII absorber so that it would
not be visible to us directly. If this is the case, the radiation emitted
from the inner hot gas, including resonantly scattered FeXXV and FeXXVI
lines, is also scattered into our line of sight by the cold reflecting
matter. The line emission would be Compton scattered and then end up
having line energies shifted lower as observed. Scattered continuum
radiation would have the same shape as the cold reflection continuum.

We do however know that the optical/UV mirror (see e.g., Miller, Goodrich
\& Mathews 1991) is located outside the torus and is at least directly
visible to us, so we should see some X-ray continuum scattered from it.
(It is unlikely that this mirror is sufficiently ionized to produce the
high ionization X-ray iron lines.) This may reconcile the small value of
$f_{\rm W}$. Importantly, the EW with respect to the warm reflection
continuum, discussed before (e.g., Table 2, Sections 2.2 and 3.1), is now
meaningless since the spatial orgin of the lines and continuum is not the
same. Uncertainties in the geometry of the inner nucleus and the mirrors
make it difficult to estimate the expected luminosities from each
component.

A further possibility, which can increase the predicted EW of the lines
from highly ionized iron, is that there is velocity shear in the warm
mirror. As discussed in Section 3.1, the EW decreases as the column
density of the mirror increases as the line is absorbed. This assumes
that the scattering medium is all at the same velocity. If, however there
is shear such that the velocity doppler-shifts the line by more than its
spectral width (natural and thermal) over a typical mean free path,
corresponding to a column of $\sim10^{21}$\psqcm, then much of the line
is not resonantly absorbed in other parts of the mirror. A larger EW
then arises from greater optical depths than assumed in Section 3.1 (see
Fig.~9 in Matt et al 1996 for an indication of the effect). For a mirror
gas temperature of about $10^6\K$, the velocity shear needs to exceed
about $50$\kmps$(10^{21}{\rm\thinspace cm}^{_2})^{-1}$ for this to be
significant. The upper limit on the width of the line ($<80$~eV; Table 1)
limits the extent of this effect. In this case the intrinsic luminosity
can be less than we estimated in Section 3.1.

Higher resolution spectroscopy resolving the individual line shapes will
test and discriminate between these hypothesis. Given the complex
geometry and the possible complexities in the formation of the iron and
other X-ray lines, we expect it to reveal that NGC~1068 has a rich and
interesting spectrum.

\section*{ACKNOWLEDGEMENTS}

We thank all the member of the ASCA PV team for maintenance and operation
of the satellite and Y. Taniguchi for useful discussion. The data
analysis are carried out using FTOOLs and XSPEC provided by the Guest
Observer Facility of NASA/Goddard Space Flight Center. ACF and KI thank
the Royal Society and PPARC, respectively, for support.

\end{document}